\documentclass[12pt]{iopart}
\usepackage{amssymb}
\usepackage{epsfig}

\begin{document}

\title[Ultra-low threshold CW Triply Resonant OPO in the near IR using
PPLN]{Ultra-low threshold CW Triply Resonant OPO in the near infrared
  using  Periodically Poled Lithium Niobate}
\author{M. Martinelli
  , K.S. Zhang, T. Coudreau\footnote[1]{To whom correspondence should
    be addressed (coudreau@spectro.jussieu.fr)}, A. Ma\^{\i}tre,
  C. Fabre}
\address{Laboratoire Kastler Brossel \\
Universit\'{e} Pierre et Marie Curie, case 74 \\
75252 Paris cedex 05 France}
\date{\today}
\maketitle
\begin{abstract}
We have operated a CW triply resonant OPO using a PPLN crystal pumped
by a Nd:YAG laser at 1.06~$\mu m$ and generating signal and idler
modes in the 2-2.3~$\mu m$ range. The OPO was operated
stably in single mode operation over large periods of time with a pump
threshold as low as 500~$\mu W$.
\end{abstract}

\section{Introduction} 

Periodically poled materials have opened the possibility of designing
more efficient Optical Parametric Oscillators. In particular, they provide
large nonlinearities in frequency ranges in which usual crystals using
birefringence for phase matching were not very efficient. This was the case
for example for the parametric generation of signal and idler fields in the
2 - 2.3~$\mu m$ region using a Nd:YAG laser as a pump, where only critically
phase matched crystals were available : parametric oscillation was reported
in this region only with high peak power pulsed Nd:YAG lasers
\cite{OPO2micron, Lin}, or in the CW regime using ultra-low loss
devices with internal
reflection~\cite{cwOPO2micron}. The availability of periodically
poled LiNbO$_3$ crystals has opened the way to very promising new
devices in this frequency range, which is of particular interest for
molecular spectroscopy and LIDAR applications \cite{lidar}. Let us
quote in particular a pulsed singly resonant OPO~
\cite{Myers,Bosenberg} with broad tunability and significant output
power, and a pump-enhanced singly resonant single mode CW OPO, with a minimum
threshold well below 1~$mW$~\cite{boller}. We describe here
another device based on a PPLN crystal, namely a Nd:YAG laser pumped,
signal-idler-pump triply resonant CW OPO working in the region 2-2.3
$\mu m$, which has a threshold in the $mW$ range.

\section{Description of the OPO}

The nonlinear crystal is a $0.5~mm \times 19~mm \times 12~mm$ PPLN
from Crystal Technology, with 8 different poling periods ranging from
30~$\mu m$ to 31.2~$\mu m$. It is inserted in a 65~mm long linear
optical cavity closed by two mirrors having a radius of curvature of
30~$mm$. This ensures mode waist sizes at the center of the crystal of
36~$\mu m$ at 1.06~$\mu m$ and 51~$\mu m$ at 2~$\mu m$, large enough
to prevent any losses when the beam crosses the 500~$\mu m$ high end
faces of the crystal.

The pump source is made of a CW monolithic diode-pumped Nd:YAG laser
(Lightwave 126-1064-700) spatially and frequency filtered using a
resonant Fabry-Perot cavity. This filtering was necessary to ensure
that the pump laser is shot-noise limited above 5~$MHz$ for quantum
noise reduction measurements \cite{squeezing} but the OPO could also
be operated without it. A pair of lenses ($f=300~mm$ and $f=60~mm$)
assures a mode matching of the pump beam to the OPO cavity of 97 \%.
The OPO cavity input mirror has an intensity reflection coefficient of
87\% at 1.06~$\mu m$ and 99.8\% in the range 2-2.2~$\mu m$. The output
mirror has a reflection coefficient of 99.8 at 1.06$~\mu m$ and
roughly 99\% in the range 2-2.2~$\mu m$. The end faces of the crystal
are anti-reflection coated at both wavelengths (residual reflection
coefficient : 0.6\% at 1.06~$\mu m$ and 0.5\% in the range 2-2.2~$\mu
m$). This gives theoretical cavity finesses of roughly 37 at the pump
frequency and 200 at 2.2~$\mu m$. The experimental measurement of the
finesse at 1.06~$\mu m$ in conditions where OPO operation was not
present gives a finesse in good agreement with the theoretical result.
It is important to note that our OPO was not designed to maximize the
signal-idler output but rather to optimize the pump squeezing
experiment \cite{Katsuyuki}, which requires low pump to signal-idler
energy transfers.

\section{OPO oscillation characteristics}

\begin{figure}[tbp]
\centerline{\epsfig{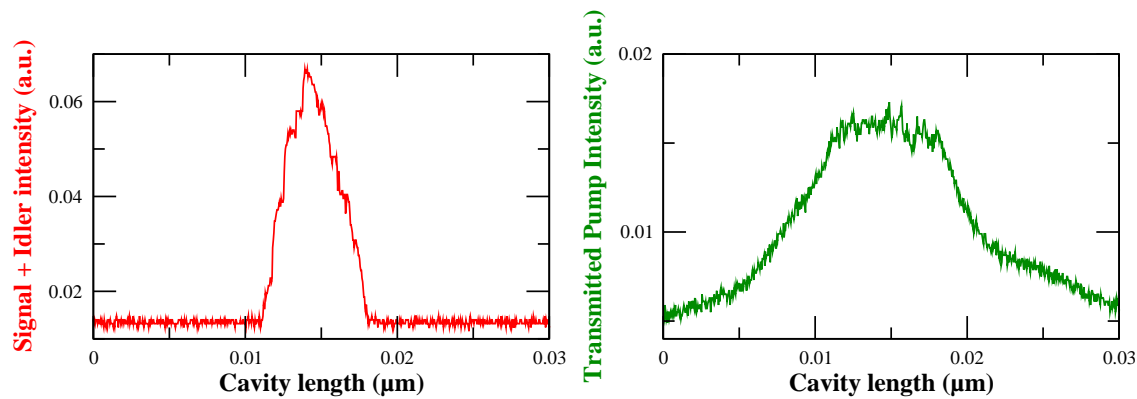}}
\caption{Signal plus idler (left) and transmitted pump (right)
  intensities as the cavity length is scanned in the vicinity of a
  pump resonance.} 
\label{balayage}
\end{figure}

Figure \ref{balayage} shows a typical recording of the intensity of
light transmitted through the output mirror respectively at 1.06~$\mu
m$ and around 2~$\mu m$, as a function of time when the cavity length
is modified via a piezoelectric ceramic. Generation of infrared light
and a strong pump depletion are observed when the cavity length is
close to a resonance with the pump frequency. When one uses the
31.1~$\mu m$ period poled region, and a crystal heated at 160$^\circ$,
the oscillation threshold was measured to be below 500~$\mu W$. This
ultra-low threshold was limited in cavity length by the pump resonance
and was obtained within 0.5 $K$ around the degeneracy temperature. We
have also obtained OPO oscillation with a period of 30.95 $\mu m$. In
that case, close to the corresponding frequency degeneracy temperature
($\approx 190^\circ C$), the threshold was similar. The other paths
had very large frequency degeneracy temperatures (above 220$^\circ$)
which made their use unpractical.

\begin{figure}[tbp]
\centerline{\epsfig{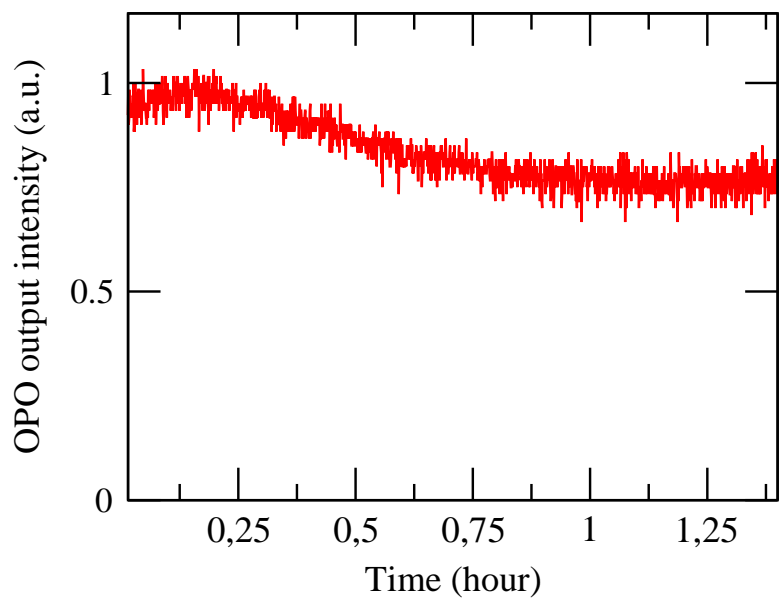}}
\caption{OPO output as a function of time}
\label{stabilite}
\end{figure}

The infrared signal can be used in a feedback loop to stabilize the
intensity of the OPO. The error signal is obtained using a lock-in
amplifier and a modulation of the cavity length at a frequency of
1~$kHz$ . With this technique, the output power residual fluctuations 
are of the order of a few \%. We have been able to keep the cavity
locked for a period over 1 hour (fig. \ref{stabilite}). The power
drift was mostly due to a slow temperature drift of the crystal.

\begin{figure}[tbp]
\centerline{\epsfig{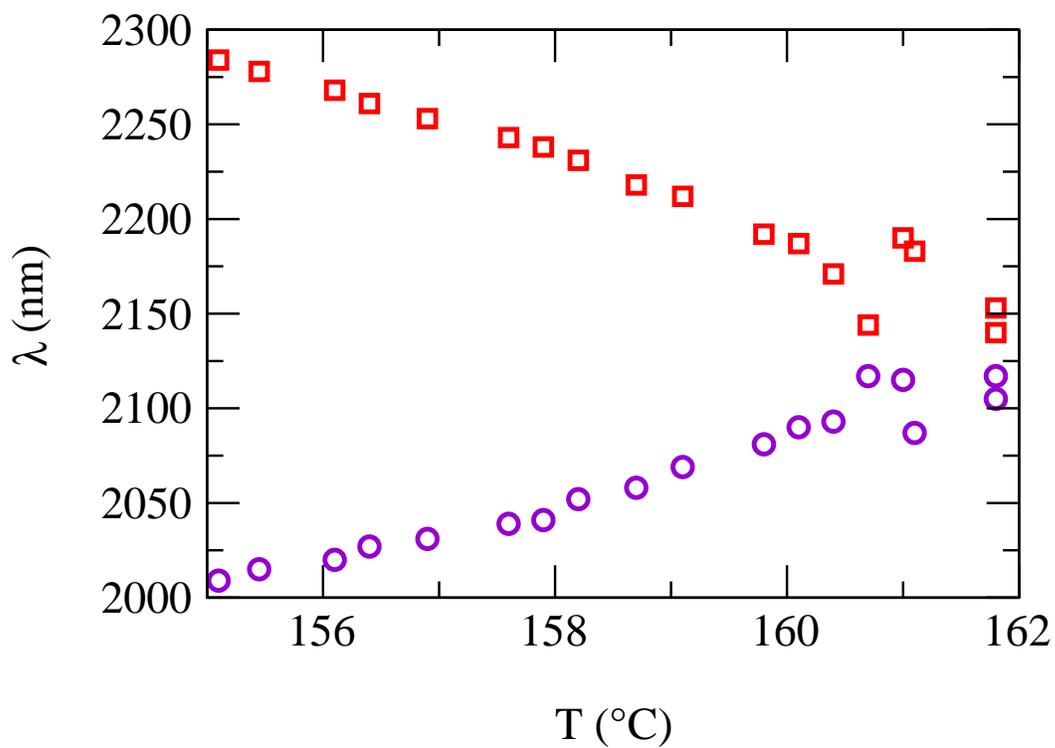}}
\caption{Signal ($\circ$) and idler ($\square$) measured wavelengths
  versus crystal temperature}
\label{ldo}
\end{figure}

Figure \ref{ldo} gives the measured wavelengths of the signal and
idler beams as a 
function of the crystal temperature when the cavity length is dithered close
to a pump-cavity resonance. The well known parabolic shape,
characteristics of a type I crystal is obtained, with some extra
points lying outside the parabola, especially close to the degeneracy
point. The same kind of behavior has been already observed in CW OPOs
\cite{Eck90}. The explanation is simple : the parabola gives the locus
of perfect phase matching as a function of temperature. The OPO may
oscillate at wavelengths that lie in 
the neighborhood of this curve, within a range given by the gain bandwidth,
and at specific values imposed by the signal idler double resonance
conditions. As already observed \cite{Eck90}, this range is especially broad
in the vicinity of the degeneracy point. 
As the cavity
length is scanned across pump resonance, the signal and idler modes
frequencies emitted change on a very short range while, for a given
length, only one pair of signal/idler modes is generated. A precise
check of the single mode character of our OPO was made using a high
finesse Fabry-Perot around 2~$\mu m$ showing mode hops as the OPO
cavity length was scanned, the beat note frequency change between mode hops
corresponding to the free spectral range of the OPO cavity. It is
important to note that close to degeneracy, as long as the pump
intensity is sufficient (\emph{i.e.} within a pump resonance),
there is  oscillation at any cavity length. This is in contrast with
the case of the type-II OPO where oscillation takes place only in
small cavity lengths intervals around lengths corresponding to the
exact double resonance condition.

\begin{figure}[tbp]
\centerline{\epsfig{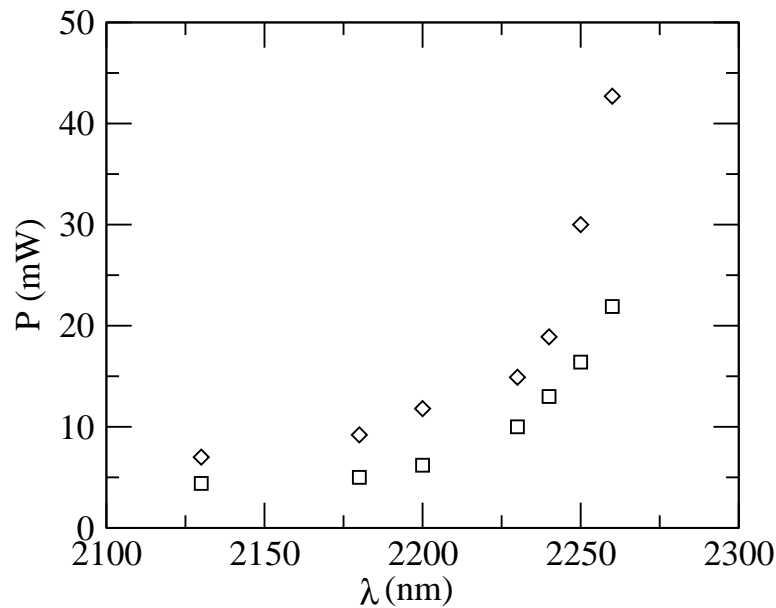}}
\caption{Experimental($\diamond$) and theoretical($\square $)values of the
OPO threshold as a function of the signal wavelength}
\label{seuil}
\end{figure}

The upper curve of Figure \ref{seuil} gives the experimental value of the OPO
oscillation threshold, which is equal to 5~$mW$ close to degeneracy,
and increases to 40~$mW$ for the signal-idler couple (2.25$\mu
$m-2.00$\mu $m ).  The lower curve gives the theoretical expectation,
corresponding to the formula \cite{sozopol}:
\begin{equation}
P_{threshold}=\frac{\left( T_p+L_p\right) ^2 \left(
T_s+L_s\right) \left( T_i+L_i\right) }{64 T_{p} \chi}
\end{equation}
where $T_{p},\, L_{p},\, T_{s},\, L_{s},\, T_{i},\, L_{i}$ are the
intensity transmission and loss coefficients for the pump signal and
idler modes respectively, and $\chi$ the effective nonlinear
coefficient taking into account in particular 
the three interacting Gaussian modes geometrical overlap. The ratio
between the theory and the experiment is roughly 
constant, of the order of 0.4, which can be considered as
satisfactory, in view of the experimental calibration uncertainties,
and of the various simplifying assumptions leading to the previous
formula (mainly the fact that there is no phase shift accumulated
between the two nonlinear interactions occurring in one round trip in
the linear cavity which is well known to increase threshold
\cite{Debuisschert}). The increase in the threshold value when one is 
far from the degeneracy point arises from the higher transmission and
loss factor at these wavelengths, due to the fact that the mirror
reflection curves have been designed to optimize the OPO close to
degeneracy. 

\begin{figure}[tbp]
\centerline{\epsfig{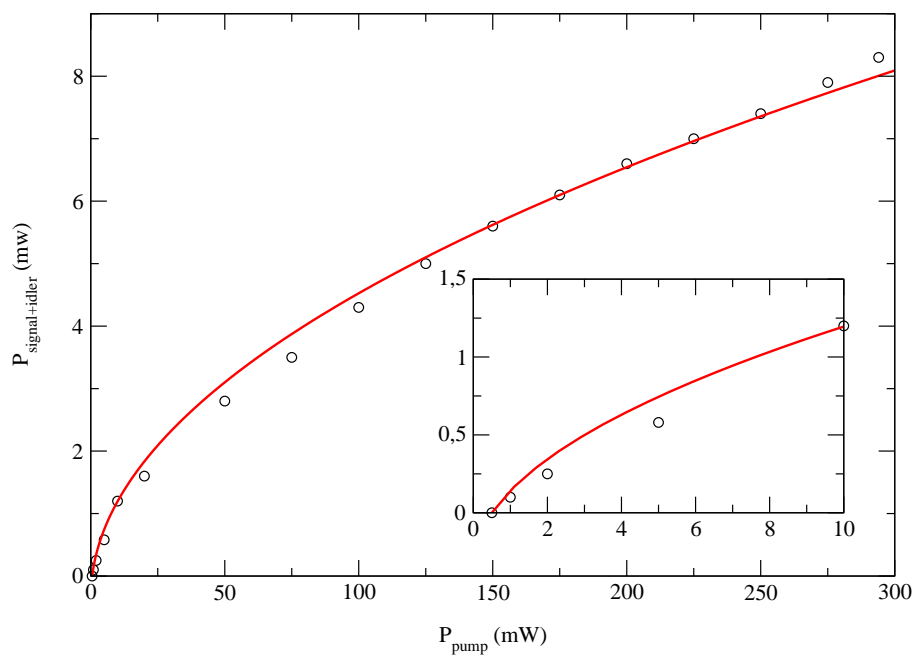}}
\caption{signal+idler modes output power as a function of pump power in the
nearly degenerate configuration (thin line with circles, experiment,
thick line theory). The inset shows the behaviour of the OPO at low
pump power and the measured threshold of 500 $\mu W$. \label{pout}}
\end{figure}

This minimum threshold of 5 $mW$ was observed in a first series of
experiments during which a spectrometer in the 2~$\mu m$ 
range was available. We have thus obtained the gross dependence of the
threshold with temperature. In a second series of experiments, we have
improved the set-up by adding a filtering cavity for the pump in order
to improve mode-matching and thus the oscillation threshold. Figure
\ref{pout} shows the variation of the 
output power as a function of pump power close to degeneracy in this
improved version of the setup. A rather
satisfactory agreement between the experimental data and the theoretical
formula (in which the experimental value of the threshold has been
used) is obtained :

\begin{equation}
P_{signal}^{out}=\hbar \omega _{signal}\frac{T_{s}\left( T_{i}+L_{i}\right)
\left( T_{p}+L_{p}\right) }{16\left| \chi ^{\prime }\right| ^{2}}\left( 
\sqrt{\frac{P_{pump}}{P_{threshold}}}-1\right) 
\end{equation}

The maximum output power for a 300~$mW$ pump and in the conditions of minimum
threshold is of 8~$mW$. The low
conversion efficiency to the signal and idler modes is of course due to the
fact that the transmission of the output mirror is smaller than the losses
(mainly on the anti-reflection coatings) for these modes. This configuration
has been chosen to minimize the threshold, and not to optimize the
conversion efficiency. 

\section{Conclusion}

We have demonstrated stable operation of a triply resonant OPO with an
ultra-low threshold below 1~$mW$ close to degeneracy  using
periodically-poled Lithium Niobate pumped by a Nd:YAG laser. Farther
from degeneracy, the signal-idler frequencies follow the expected
parabola-shaped  curve. The ultra low threshold obtained allows
operation well above threshold even for moderate pump powers far from
the saturation of photodiodes. This has enabled
us to demonstrate quantum noise reduction on the pump laser using
periodically-poled materials \cite{squeezing}. 

\ack 
M.M. wishes to thank Coordena\c c$\tilde\mathrm a$o de Aperfei\c
coamento de Pessoal de N\' \i vel Superior (CAPES-BR) for
funding. This research was performed in the framework of the EC ESPRIT
contract  ACQUIRE 20029.


\begin{thebibliography}{99}
\bibitem{OPO2micron}  R.\ Herbst, R.\ Fleming, R.\ Byer Applied Physics
Letters \textbf{25}, 520 (1974)

\bibitem{Lin}  J.\ Lin, J.\ Montgomery, Optics Commun. \textbf{75}, 315
(1990)

\bibitem{cwOPO2micron}  D.\ Serkland, R.\ Eckardt, R.\ Byer, Optics Letters 
\textbf{19}, 1046 (1994)

\bibitem{lidar}  R.\ Frehlich, Journal of Atmospheric and Oceanic
Technology, \textbf{12} (1995)

\bibitem{Myers}  L.\ Myers, R.\ Eckardt, M.\ Fejer, R.\ Byer, W.\ Bosenberg,
Optics Letters \textbf{21}, 591 (1996)

\bibitem{Bosenberg}  W.\ Bosenberg, A.\ Drobshoff, J.\ Alexander, L.\ Myers,
R.\ Byer, Optics Letters \textbf{21}, 1336 (1996)

\bibitem{boller}  K.-J. Boller,, M.E. Klein, D.-H. Lee, P. Gross,
  H. Ridderbusch, M.A. Tremont, A. Robertson, J.-P. Meyn,
  R. Wallenstein, CLEO oral communication, Nice (sept. 200) 

\bibitem{squeezing} K.S. Zhang, T. Coudreau, M. Martinelli,
  A. Ma\^{\i}tre, C. Fabre, to be published in Phys. Rev. \textbf{A}

\bibitem{Katsuyuki} K. Kasai, JiangRui Gao, C. Fabre,
  Europhys. Lett. \textbf{40}, 25 (1997) 

\bibitem{Eck90}  R. Eckardt, C.\ Nabor, W.\ Kozlovsky, R.\ Byer J.\ Opt.\
Soc.\ Am.\ \textbf{B8}, 646 (1991)

\bibitem{sozopol} C. Fabre, in \textit{Advanced Photonic with
    Second-order Optically Nonlinear Processes}, A.D. Boardman
  \textit{et al.} (eds.), Kluwer Academic Publishers (1999)

\bibitem{Debuisschert}  T.\ Debuisschert, A.\ Sizmann, E.\ Giacobino, C.\
Fabre J.\ Opt.\ Soc.\ Am.\ \textbf{B10}, 1668 (1993)



\end{thebibliography}
\end{document}